\documentclass[aps,prd,reprint,nofootinbib,preprintnumbers]{revtex4-1}

\usepackage[T1]{fontenc}
\usepackage{newtxtext}
\usepackage[varg]{newtxmath}

\usepackage{titlesec}

\newcommand{\mysectionnumbering}{\thesection.~}

\titleformat{\section}{\bfseries\center\uppercase}{\mysectionnumbering}{0em}{}
\titleformat{\subsection}{\bfseries\center}{}{0.1em}{\thesubsection.~}
\titleformat{\subsubsection}{\bfseries\itshape\center}{}{0.1em}{\thesubsubsection.~}
\titleformat{\paragraph}[runin]{\itshape}{}{0em}{(\theparagraph)\hspace{0.5em}}[.]

\titlespacing{\section}{0pt}{2em plus 0.1em minus 0.1em}{0.7em}
\titlespacing{\subsection}{0pt}{1.5em}{0.5em}
\titlespacing{\subsubsection}{0pt}{1.5em}{0.5em}

\let\oldappendix\appendix

\usepackage{etoolbox}
\makeatletter
\renewcommand{\appendix}{\@ifstar\oneappendix\manyappendices}
\makeatother

\newcommand{\defappsec}{%
	\let\oldsection\section
	\renewcommand{\section}[1]{%
	\ifstrempty{##1}
		{\oldsection{}}
		{\oldsection{:~{##1}}}
	}
}

\newcommand{\oneappendix}{%
\oldappendix*
\defappsec
\renewcommand{\mysectionnumbering}{\MakeUppercase{Appendix}}
\renewcommand\theequation{\Alph{section}\arabic{equation}}
}
\newcommand{\manyappendices}{%
\oldappendix
\defappsec
\renewcommand{\mysectionnumbering}{\MakeUppercase{Appendix}~\thesection}
\renewcommand\theequation{\Alph{section}\arabic{equation}}
}

\makeatletter
\renewcommand\@makefntext[1]{%
\noindent{\hspace{1em}}{\@makefnmark}#1}
\makeatother

\makeatletter
\renewcommand\@makefnmark{\hbox{\@textsuperscript{\normalfont\color{black}\@thefnmark}}}
\makeatother

\renewcommand{\footnoterule}{%
  \kern -3pt
  \hrule width 1.2cm
  \kern 4pt
}

\renewcommand{\dot}[1]{\overset{\bm .}{#1}}

\newcommand{\eqskip}{{\newpage~\vskip-2.5\baselineskip}}

\DeclareSymbolFont{cmodern}{OML}{cmm}{m}{i}
\DeclareMathSymbol{\partial}{0}{cmodern}{64}
\DeclareMathSymbol{g}{\mathalpha}{cmodern}{103}

\usepackage{esint}

\usepackage[dvipsnames]{xcolor}
\usepackage[colorlinks=true,allcolors=Blue,breaklinks=true]{hyperref}

\usepackage{enumitem}
\setlist[enumerate]{label={(\arabic*)},leftmargin=2.7em,topsep=1pt,itemsep=2pt,parsep=0em}

\usepackage{bm,xspace}

\newcommand{\bmf}[1]{\mathbf{#1}}
\newcommand{\dx}{\textup{d}}
\newcommand{\w}{\omega}
\newcommand{\avg}[1]{{\langle #1 \rangle}}
\newcommand{\K}{N}
\newcommand{\Y}{\mathcal{Y}}
\newcommand{\M}{\mathcal{M}}
\newcommand{\J}{\mathcal{J}}

\DeclareMathAlphabet{\mathcal}{OMS}{cmsy}{m}{n}

\begin{document}

\title{Superradiant scattering by a black hole binary}

\author{Leong Khim \surname{Wong}}

\affiliation{DAMTP, Centre for Mathematical Sciences, University of Cambridge,
Wilberforce Road, Cambridge CB3 0WA, United Kingdom}

\date{24 August 2019}

\begin{abstract}
I present evidence of a novel guise of superradiance that arises in black hole binary spacetimes. Given the right initial conditions, a wave will be amplified as it scatters off the binary. This process, which extracts energy from the orbital motion, is driven by absorption across the horizons and is most pronounced when the individual black holes are not spinning. Focusing on real scalar fields, I demonstrate how modern effective field theory (EFT) techniques enable the computation of the superradiant amplification factor analytically when there exist large separations of scales. Although exploiting these hierarchies inevitably means that the amplification factor is always negligible (it is never larger than about one part in $10^{10}$) in the EFT's regime of validity, this work has interesting theoretical implications for our understanding of general relativity and lays the groundwork for future studies on superradiant phenomena in binary systems.
\end{abstract}

\maketitle

\section{Introduction}
Superradiance is an intriguing phenomenon wherein waves scattering off a rotating body are amplified. It is perhaps most widely known to occur around Kerr black holes \cite{Starobinsky:1973aij,*Starobinsky:1973aij_v2,Starobinskil:1974nkd,*Starobinskil:1974nkd_v2,Teukolsky:1972my,Teukolsky:1974yv} and, in this context, can be viewed as the wave analog of the Penrose process \cite{Penrose:1969pc,*Penrose:1969pc_v2,Penrose:1971uk}. However, superradiance is not in itself intrinsically tied to the existence of an ergosphere but is in fact a far more generic process \cite{[{For reviews, see }]Bekenstein:1998nt,*Brito:2015oca}. Case in point: In his seminal papers on the subject \cite{Zeldovich:1971,*Zeldovich:1971_v2,Zeldovich:1972,*Zeldovich:1972_v2}, Zel'dovich describes the amplification of electromagnetic waves by a conducting cylinder; while more recently, Torres \emph{et al.}~reported the first laboratory observation of superradiance in water waves scattered by a draining vortex \cite{Torres:2016iee}. Superradiant scattering by rotating stars has also recently been studied~\cite{Richartz:2013hza,Cardoso:2015zqa,Cardoso:2017kgn}.

Diverse as these systems are, they can all be distilled down to the same two essential ingredients: a reservoir of energy that can be extracted and a channel for dissipation \cite{Richartz:2009mi}. With these in mind, it is natural to expect that a black hole \emph{binary} should also exhibit superradiance. Just like in the single black hole case, dissipation is provided by absorption across the horizons, but now the predominant source of rotational energy is the binary's orbital motion rather than the spins of its constituents. Consequently, such binaries ought to amplify waves even when the individual black holes are not spinning. The goal of this work is to substantiate these expectations by presenting the first-ever analytic calculations for this process.

These calculations may be important for astrophysics and gravitational-wave science, given how common binaries are in our Universe, but they also present an interesting theoretical challenge. The traditional approach to studying (single) black hole superradiance involves solving a wavelike equation on a fixed Kerr background. While highly successful, extending this approach to more complicated scenarios is prohibitively difficult. No exact solution to the Einstein equations describing an inspiraling binary is known, but even if one were, the corresponding wavelike equation on this spacetime is likely intractable. The nonlinearities that encumber this approach are mostly inessential to the problem, however, and the breakthrough that has paved the way for this present work is an effective field theory (EFT) formalism for extended objects \cite{Goldberger:2004jt,Goldberger:2005cd,Goldberger:2009qd,Porto:2016pyg}. By exploiting an inherent separation of scales, a black hole interacting with a low-frequency, long-wavelength field can be approximated as a point particle furnished with dynamical operators that capture finite-size effects.

Building on this formalism, Endlich and Penco \cite{Endlich:2016jgc} have shown how quantum field theoretic techniques can be used to study superradiance by converting the problem into one of computing S-matrix elements. Their approach has hitherto been limited to systems that are stationary and (nearly) spherically symmetric, however, while we will find that it is the absence of these symmetries that makes binary systems especially interesting. In striving to extend their approach to the latter case, we have found it more instructive to forego S-matrix elements in favor of working directly with equations of motion.

This paper proceeds as follows: We begin in Sec.~\ref{sec:eft} by discussing the relevant degrees of freedom (DOF) in the EFT and how they interact. Subsequently, the effective equation of motion for a wave propagating on a fixed background binary spacetime is derived. In Sec.~\ref{sec:scattering}, we then solve this equation perturbatively to obtain the amplification factor. (Some of the more technical details are relegated to an Appendix.) The mechanism by which energy is extracted from the binary is also clarified at this stage. Finally, a summary and a brief discussion on avenues for future work are presented in Sec.~\ref{sec:discussion}. For simplicity, throughout this paper we concentrate solely on the superradiance of a real scalar field, leaving higher-spin fields to the future.

\section{The effective theory}
\label{sec:eft}

Our goal here is to construct an effective theory that describes how a scalar field $\phi$ scatters off a black hole binary. Three simplifying assumptions facilitate analytic calculations:
\begin{enumerate}
	\item We will assume that the binary is in the early phase of its inspiral, during which its orbital separation $a$ is much greater than the typical size $GM$ of its constituents (let ${\hslash = c = 1}$). As a result, the characteristic orbital velocity ${v \sim (GM/a)^{1/2}}$ is small. This hierarchy permits use of the aforementioned point-particle approximation, wherein the black holes are replaced by worldlines of some effective center-of-energy coordinate.
	
	\item Let us further assume that the scalar has a wavelength ${\lambda \gg a}$. This regime permits a coarse-grained description of the system, in which the binary itself looks like an effective point particle from the field's perspective. 
	
	\item Finally, we neglect the possibility of resonantly exciting individual black holes \cite{Simone:1991wn} by concentrating on low-energy fields varying on a timescale $\w^{-1}$ much longer than the black holes' light-crossing times.%
\footnote{Note that ${ GM\w \sim v^2 (a/\lambda) v_p }$, where $v_p \geq 1$ is the phase velocity of the wave. For massless fields (${v_p = 1}$), we automatically have ${GM\w \ll 1}$ if the first two assumptions are satisfied, but for massive fields this is an independent assumption.}
\end{enumerate}

These three assumptions establish an EFT organized as an expansion in three small parameters. Working perturbatively in powers of $v$ yields the typical post-Newtonian (PN) expansion \cite{Blanchet:2013haa}, which is here supplemented by the additional parameters $a/\lambda$ and $GM\w$ that characterize the interactions between the scalar and the binary. For this preliminary study, let us restrict our attention to leading order in these parameters. It then suffices to consider a Newtonian binary with total mass $M=M_1+M_2$ and orbital frequency~$\Omega$. We will further assume a circular orbit for added simplicity. Place its barycenter at the origin and orient the orbital angular momentum vector to be along the positive $z$~axis. The constituent black holes, labeled by $\K \in \{1,2\}$, travel along the worldlines $\bmf z_\K^i(t) = r_\K R^i{}_j(t) \bmf d^j$. Their distances from the origin are given by $r_1 = a M_2/M$ and $r_2 = -a M_1/M$, $\bmf d$ is a unit vector in the $z=0$ plane that specifies their initial positions, and $R^i{}_j(t)$ is the appropriate rotation matrix.

\subsection{Interaction terms}

Appropriate boundary conditions for $\phi$ must be specified at the positions of the worldlines $\bmf z_\K(t)$ to account for absorption. This is achieved in the EFT formalism by introducing interaction terms that couple the scalar to a set of microscopic DOF $q^L_\K(t)$ localized on the worldlines \cite{Goldberger:2005cd,Endlich:2016jgc,WDG}:
\begin{equation}
S_\text{int} = \int\dx t \sum_{\K=1}^2\sum_{\ell=0}^\infty 
q^L_\K(t) \partial_L \phi{\bm(} t,\bmf z_\K(t) {\bm)}.
\label{eq:S_int}
\end{equation}
For each $\K$, the sum over $\ell$ decomposes the scalar field into distinct spherical harmonic modes centered on the black hole.%
\footnote{We use conventional multi-index notation \cite{Blanchet:2013haa}: A tensor with $\ell$ spatial indices is written as $q^L \equiv q^{i_1 \dots i_\ell}$, whereas $\ell$ factors of a vector are written as $\bmf z^L \equiv \bmf z^{i_1} \dots \bmf z^{i_\ell}$, and similarly, $\partial_L \equiv \partial_{i_1} \dots \partial_{i_\ell}$.} 
Accordingly, the composite operators $q^L_\K(t)$, which are taken to be symmetric and trace free (STF), correspond to the black hole's dynamical multipole moments. These can exchange energy and momentum with the scalar, thus modeling the absorptive property of the black hole's horizon.

As ${ |\bmf z|\sim a \ll \lambda }$ by assumption, we can Taylor expand $\phi$ about the origin and organize terms such that \eqref{eq:S_int} now reads
\begin{equation}
S_\text{int} = \int\dx t \sum_{\ell=0}^\infty
O^L(t) \partial_L
\phi(t,\bmf 0).
\label{eq:S_int_O} 
\end{equation}
This form is better suited to our purposes: The new sum over $\ell$ decomposes the scalar into spherical harmonics centered on the origin and the interactions have been regrouped into a set of STF operators $O^L$, which correspond to the dynamical multipole moments of the binary. At leading order,
\begin{equation}
O^L(t) =
\frac{1}{\ell!}\sum_\K q_\K(t) \bmf z_\K^\avg{L}(t),
\label{eq:O}
\end{equation}
where angled brackets denote the STF projection of a tensor.

Notice that all operators $q_\K^L$ with $\ell \geq 1$ have been discarded. Power counting reveals that their correlation functions scale as $\avg{q^L q^L}/\avg{q q} \sim (\text{length})^{2\ell}$ \cite{WDG}, and since these operators are related to short-distance physics near a black hole's horizon, the appropriate length scale must be $GM$. Consequently, a term of the form ${q^L \bmf z^{i_{\ell+1}} \dots \bmf z^{i_n}}$ is suppressed relative to ${q^{L-1}\bmf z^{i_\ell}\dots \bmf z^{i_n}}$ by one power of ${GM/a \sim v^2}$; thus only the $\ell=0$ operator $q_\K(t)$ is needed at leading order. Physically, this scaling analysis tells us that a black hole's absorption cross section is $s$-wave dominated in the low-frequency limit \cite{FHM}.

To see why traces were neglected, consider the quadrupole operator $O^{ij}$ as an example. If kept, its trace would contribute a term of the form $O^k{}_k \partial^i \partial_i \phi$. Because a scalar of mass $\mu$ satisfies the Klein-Gordon equation in the absence of the binary, this interaction term is unchanged at leading PN order were we to replace $\partial^i \partial_i \phi \to (\partial_t^2 + \mu^2)\phi$. (On a technical note, this replacement is achieved by a field redefinition \cite{Georgi:1991ch,Goldberger:2007hy}.) The new term $O^k{}_k (\partial_t^2 + \mu^2)\phi$ no longer depends on spatial derivatives and so can be absorbed into a redefinition of $O(t)$. Power counting tells us $|O^k{}_k (\partial_t^2 + \mu^2)\phi|/|O(t)\phi| \sim (a/\lambda)^2$; hence, this is a subleading correction. The pattern extends to all multipoles: Traces of $O^L$ can always be converted into subleading corrections to lower-multipole operators.

\subsection{Equations of motion}
Extremizing the full action with respect to $\phi$ yields
\begin{equation}
(\Box - \mu^2)\phi = - \sum_{\ell=0}^\infty (-1)^\ell O^L(t) \partial_L \delta^{(3)}(\bmf x).
\label{eq:eom_1}
\end{equation}
The rhs depends on $q_\K(t)$, which is a dynamical variable in its own right, meaning it must satisfy its own equation of motion. However, as this operator is a microscopic DOF related to short-distance physics near the horizon, the low-energy (long-distance) EFT supplies no specific details about what this equation might be, so additional input is required.

On general grounds, we expect $q_\K$ to satisfy an equation of the form
\begin{equation}
\label{eq:eom_q}
\mathcal D q_N(t) = - \sum_{\ell=0}^\infty \frac{1}{\ell!} \bmf z_\K^\avg{L}(t) \partial_L \phi(t,\bmf 0),
\end{equation}
where $\mathcal D$ is some (possibly nonlinear) differential operator, while the source term on the rhs comes from extremizing $S_\text{int}$ with respect to $q_\K$. No-hair theorems tell us that isolated black holes cannot support their own permanent scalar moments \cite{Bekenstein:1971hc,Hawking:1972qk,Sotiriou:2011dz}; hence, the solution to \eqref{eq:eom_q} must vanish unless there is an external source present. We therefore expect the leading-order solution to be given by linear response theory:
\begin{equation}
q_\K(t) = \int\dx t' \chi_{R,\K}(t-t')
\left( \sum_{\ell=0}^\infty \frac{1}{\ell!} \bmf z_\K^\avg{L}(t') \partial_L \phi(t',\bmf 0) \right).
\label{eq:q}
\end{equation}
Any nonlinearities in the response can be neglected when the scalar's amplitude is not too large. Because the spacetime of an isolated black hole is stationary, the retarded Green's function $\chi_R$ is necessarily invariant under time translations. (To be precise, $\chi_R$ is invariant under translations of the proper time $\tau$ along the worldline, but $\tau \approx t$ in the Newtonian limit.)

To determine $\chi_R$, we use the fact that, in the low-frequency limit, its Fourier transform admits a Taylor expansion in powers of $GM\w$ \cite{Endlich:2016jgc}. The coefficients of this expansion are then fixed by matching them to observables calculated using the full Kerr solution. For instance, one finds by calculating the accretion rate of a scalar field onto a single black hole at rest that ${\tilde\chi_{R,\K}(\w) = iA_\K \w}$ at leading order \cite{WDG}, where $A_\K$ is the area of the $\K$th black hole. The power of this matching procedure is that the values of these (Wilsonian) coefficients are universal and so once determined can be applied to more complicated scenarios, like in the present context.

Combining \eqref{eq:O}, \eqref{eq:eom_1}, and \eqref{eq:q} leaves us with a linear, homogeneous equation of motion for $\phi$ to solve.

\section{Superradiant scattering}
\label{sec:scattering}

In effect, this EFT is exploiting the hierarchy ${a \ll \lambda}$ to zoom out on the binary and replace it by a set of interaction terms localized at the origin. To leading PN order, the spacetime everywhere else is flat; hence, the  general solution for $r > 0$ is
\begin{equation}
\label{eq:phi_sol_general}
\phi(x) = \sum_{\ell,m}\int_\w e^{-i\w t}
[ \mathcal I_{\w\ell m} h^-_\ell(kr) + \mathcal R_{\w\ell m} h^+_\ell(kr) ]
Y_{\ell m}(\hat{\bmf x}),
\end{equation}
where $Y_{\ell m}(\hat{\bmf x})$ are the spherical harmonics and we write ${ \int_\w \equiv \int\dx\w/(2\pi) }$. The wave number $k$ is defined as the appropriate root of ${k^2 = \w^2 - \mu^2}$, namely
\begin{equation}
k(\w) \coloneq
\begin{cases}
i \sqrt{\mu^2 - \w^2} & \w^2 \leq \mu^2 \\
\text{sgn}(\w) \sqrt{\w^2 - \mu^2} & \w^2 \geq \mu^2.
\end{cases}
\end{equation}

The radial part of the solution~\eqref{eq:phi_sol_general} is given by the spherical Hankel functions, which have the limiting forms $ h_\ell^\pm(z) \sim i^{\mp(\ell+1)}e^{\pm iz}/z $ as ${z \to \infty}$. Accordingly, $\mathcal I_{\w\ell m}$ and $\mathcal R_{\w\ell m}$ are the amplitudes for ingoing and outgoing waves, respectively. [For real $\phi$, we must have ${\mathcal I_{\w\ell m}^* = (-1)^{\ell+m} \mathcal I_{-\w\ell-m}}$ and likewise for $\mathcal R^*_{\w\ell m}$.]

\subsection{Amplification factor}
The relationship between $\mathcal I_{\w\ell m}$ and $\mathcal R_{\w\ell m}$ is determined by the interaction terms at the origin (specified by $O^L$), which we will treat as small perturbations. In the absence of these interactions, there cannot be a net flow of energy into or out of the origin; hence, ${\mathcal I_{\w\ell m} = \mathcal R_{\w\ell m}}$ at zeroth order. Turning on the interactions, the outgoing amplitude becomes ${\mathcal R_{\w\ell m} = \mathcal I_{\w\ell m} + \mathcal A_{\w\ell m}}$. We determine $\mathcal A_{\w\ell m}$ by computing the first-order solution to \eqref{eq:eom_1}:
\begin{equation}
\label{eq:phi_sol_1}
\phi^{(1)}(x) = \int\dx^4x' G_R(x-x') \sum_{\ell=0}^\infty (-1)^\ell O^L(t') \partial_L \delta^{(3)}(\bmf x'),
\end{equation}
where $G_R$ is the retarded Klein-Gordon propagator and $O^L(t')$ is evaluated using the zeroth-order solution $\phi^{(0)}$. With some effort (details of which are presented in the Appendix), it can be shown that
\begin{equation}
\label{eq:phi_sol_1_A}
\phi^{(1)}(x) = \sum_{\ell,m}\int_\w e^{-i\w t}
\mathcal A_{\w\ell m} h^+_\ell(kr) Y_{\ell m}(\hat{\bmf x}),
\end{equation}
with
\begin{align}
\mathcal A_{\w\ell m} &=
\sum_{\ell',m'}\int_{\w'}
\frac{2 Y^*_{\ell m}(\bmf d) Y_{\ell'm'}(\bmf d) \mathcal I_{\w'\ell'm'}}{(2\ell+1)!!(2\ell'+1)!!}
 \sum_\K A_\K  r_\K^{\ell+\ell'} k^{\ell+1}
\nonumber\\ &\quad\times
(k')^{\ell'} (m'\Omega - \w')  2\pi\delta{\bm(}\w-\w' -(m-m')\Omega{\bm)}.
\label{eq:A}
\end{align}
Beyond just being a feasible way of solving \eqref{eq:eom_1}, this perturbative approach also benefits from having an intuitive physical picture: The unscattered wave $\phi^{(0)}$ induces a scalar monopole moment---or scalar ``charge''---onto each of the black holes, which are then able to radiate scalar waves.\footnote{Had we kept the higher-multipole operators $q_\K^L(t)$ with $\ell \geq 1$, the black holes would also gain higher multipole moments (but recall their effects are suppressed by powers of $v$). For more on black holes with induced scalar charges, see Refs.~\cite{WDG,Jacobson:1999vr,Horbatsch:2011ye,Healy:2011ef,Berti:2013gfa}.} This outgoing radiation $\phi^{(1)}$ is the scattered wave.

Let us now explore the physical implications of \eqref{eq:A}. This result is valid for any ingoing amplitude $\mathcal I_{\w\ell m}$, but it will be instructive to consider an ingoing wave peaked at a single frequency $\hat\w$ and composed of a single harmonic $(\hat\ell,\hat m)$.%
\footnote{That is, $\mathcal I_{\w\ell m} \propto \delta(\w-\hat\w)\delta_{\ell\hat\ell}\delta_{m\hat m} + (-1)^{\ell+m}\delta(\w+\hat\w)\delta_{\ell\hat\ell}\delta_{m,-\hat m}$.}
The delta function in \eqref{eq:A} then tells us that the binary will scatter this wave into multiple outgoing modes. For instance, the single ingoing mode $(\hat\w,1,1)$ will scatter into the outgoing modes $(\hat\w,1,1)$, $(\hat\w-\Omega,0,0)$, $(\hat\w-\Omega,2,0)$, and so on. This appearance of mode mixing is unsurprising, given that a binary is neither stationary nor axisymmetric.

Mode mixing also means that we cannot speak of the amplification of each mode individually (in contrast to single black hole superradiance), but we can still compute the total amplification factor for the wave. Integrated over all time, the total energy flux radiated off to infinity is given by $\Delta E = \Delta E_\text{out} - \Delta E_\text{in}$, where
\begin{equation}
\Delta E_\text{out} = \sum_{\ell,m}\int_\w \theta(k^2) \frac{\w}{k} |\mathcal R_{\w\ell m}|^2,
\end{equation}
$\theta(k^2)$ is the Heaviside step function, and the expression for $\Delta E_\text{in}$ is obtained by replacing $\mathcal R_{\w\ell m}$ with $\mathcal I_{\w\ell m}$. The total amplification factor is then ${Z = \Delta E/\Delta E_\text{in}}$. For a single ingoing mode $(\hat\w,\hat\ell,\hat m)$, ${Z = 2\mathcal A_{\hat\w\hat\ell\hat m}/\mathcal I_{\hat\w\hat\ell\hat m} + \mathcal O(\mathcal A^2)}$. Hence, even though a binary scatters a single ingoing mode into multiple outgoing modes, interference between the zeroth- and first-order solutions results in most of the outgoing energy being carried by the original mode. Written out explicitly,
\begin{equation}
Z =
\left|\frac{2Y_{\hat\ell\hat m}(\bmf d)}{(2\hat\ell+1)!!}\right|^2 \sum_\K A_\K r_\K^{2\hat\ell} \hat k^{2\hat\ell+1}(\hat m\Omega - \hat\w).
\label{eq:Z}
\end{equation}

Three remarks are worth making about \eqref{eq:Z}. First, the sum over $\K$ signifies that only one black hole is needed for superradiance to occur; the other member of the binary need not interact with the scalar at all. Second, superradiance occurs when $Z > 0$, which translates to two necessary conditions: The ingoing mode must satisfy the familiar inequality ${0 < \hat\w < \hat m \Omega}$ \cite{Brito:2015oca}, but additionally, ${\hat\ell + \hat m}$ must also be even. This second condition is novel to binary systems and originates from the quantity ${ |Y_{\hat\ell\hat m}(\bmf d)|^2 \equiv |Y_{\hat\ell\hat m}(\pi/2,0)|^2 }$, which vanishes when ${\hat\ell+\hat m}$ is odd \cite{DLMF}. These vanishing modes correspond to field profiles concentrated away from the ${z=0}$ plane, which makes intuitive sense---no exchange of energy can occur if the field is unappreciable in the neighborhood of the binary. Third remark: The results presented here account only for the contribution to superradiance from the binary's orbital motion, but are otherwise valid for black holes of any spin. When $v$ and $GM\w$ are small, the additional contribution from the field interacting with the spins of the individual black holes is subleading, since it arises from the Green's functions for  $q^L_\K$ with $\ell \geq 1$ \cite{Endlich:2016jgc}.

The fact that $Z \propto (a/\lambda)^{2\hat\ell}$ indicates that superradiance is most pronounced in the ${ \hat\ell = \hat m = 1 }$ mode. For a given value of $\Omega$, $Z$ is further maximized if the wave is massless with frequency ${ \w = 3\Omega/4 }$ and if both black holes are spherical. In this special case, ${Z_\text{max} = 9 \nu^2 v^8/16}$, where ${ \nu = M_1 M_2/M^2 \leq 1/4 }$ is the symmetric mass ratio. Even if we take $v \approx 0.1$ (which is already pushing the limits of validity of the EFT), we find $Z_\text{max} \approx 4 \times 10^{-10}$ at best.

\subsection{Energy balance}
Simple thermodynamic arguments provide further insight into the onset of superradiance. 
Up to 2PN order, there is no gravitational radiation and no concomitant orbital decay; hence, the spacetime for a circular binary admits the helical Killing vector $\xi = \partial_t + \Omega\partial_\varphi$. Spacetimes with this (approximate) symmetry satisfy the first law of binary black hole mechanics \cite{Friedman:2001pf,*Friedman:2001pf_E,LeTiec:2011ab}:
\begin{equation}
\dx\M - \Omega\,\dx\J = \sum_\K \frac{\kappa_\K}{8\pi G} \dx A_\K,
\label{eq:first_law}
\end{equation}
where $\kappa$ is the surface gravity, while $\M$ and $\J$ are the Arnowitt-Deser-Misner mass and angular momentum, respectively. When a single ingoing mode with ${\lambda\gg a}$ scatters off the binary, angular momentum exchange is constrained by ${ \dx\J/\dx\M = \hat m/\hat\w }$; hence, the lhs of \eqref{eq:first_law} reads ${ \hat\w^{-1} (\hat\w-\hat m\Omega)\,\dx\M }$. The rhs is positive semidefinite due to the second law (${\dx A_\K \geq 0}$); thus we easily recover the conditions ${ 0 < \hat\w < \hat m\Omega }$ and ${ \dx A_\K > 0} $ necessary for superradiance (${\dx\M < 0}$).
This reinforces the fact that absorption is still an essential ingredient, even though the spacetime we consider is time dependent. The Killing vector $\xi$ gives rise to a conserved energy-momentum current for the scalar (loosely speaking, $\xi$ generates time-translation symmetry in the frame corotating with the binary); thus at low PN orders the gravitational potential sourced by the binary plays no role in amplifying or dissipating waves.

The first law also provides another way of computing $\Delta E$. The energy gained by the binary during the scattering process is
\begin{equation}
\Delta\M = \sum_\K \int\dx t \,(\dot M_\K + \bmf F_\K \cdot \bmf v_\K),
\label{eq:E_binary}
\end{equation}
where the first term is due to absorption, while the second is the work done by the scalar. It has previously been shown \cite{WDG} in a fully relativistic setting that a background scalar ${\Phi \equiv \phi^{(0)}}$ exerts a ``fitfh force'' ${F^\mu = Q(\tau)(g^{\mu\nu} + u^\mu u^\nu)\partial_\nu\Phi}$ on a black hole due to the charge ${Q(\tau) \coloneq - A \dot\Phi{\bm(} z(\tau) {\bm)}}$ it induces.\footnote{An overdot denotes a derivative with respect to the proper time $\tau$ along the worldline and $u^\mu \equiv \dot z^\mu$.}
The fluctuation-dissipation theorem links this charge with the accretion rate, ${\dot M \equiv - Q \dot\Phi}$; hence, $F^\mu$ includes the drag due to accretion, but additionally accounts for the impact of spatial gradients of $\Phi$. Taking its nonrelativistic limit, the integrand in \eqref{eq:E_binary} can be rewritten as ${A_\K \dot\Phi (\dot\Phi - \bmf v_\K^i\partial_i \Phi)}$. Expanding $\Phi \equiv \Phi{\bm(}t,\bmf z_\K(t){\bm)}$ about the origin and using \eqref{eq:id_1} and \eqref{eq:id_2}, we indeed find ${\Delta E = -\Delta\M}$. Hence, the scalar extracts energy from the binary through the action of this fifth force $F^\mu$. This explains why ${Z \propto A_\K}$. Naively, we might guess that superradiance is suppressed for spherical black holes because of their larger absorption cross section, but the exact opposite is true---the fluctuation-dissipation theorem stipulates that $F^\mu$ is also proportional to this cross section.

\section{Discussion}
\label{sec:discussion}

Despite the inherent complexity of such systems, this paper demonstrates how the interaction between a black hole binary and an external scalar field can be understood analytically when large separations of scales are present. An EFT for this system (which is an extension of Goldberger and Rothstein's formalism \cite{Goldberger:2004jt,Goldberger:2005cd,Goldberger:2009qd,Porto:2016pyg}) was constructed via a two-step process: First, the hierarchies ${ v^2 \sim GM/a \ll 1 }$ and ${ GM\w \ll 1 }$ were used to zoom out on the black holes and replace them with point particles traveling along worldlines. Second, the hierarchy ${a/\lambda \ll 1}$ was exploited to zoom out on the binary as a whole, replacing it by a new effective point particle localized at the origin.

In this limit, an explicit calculation provides compelling evidence for the existence of a novel guise of superradiance in binary systems, which exhibits several distinct features: Mode mixing due to the absence of stationarity and axial symmetry is manifest in the equations and a new superradiance condition tied to the geometry of the problem is found. Even so, this study constitutes only the first step towards a comprehensive understanding of this process and there are multiple avenues for future work. 

For instance, our assumption of a circular orbit rendered many of the calculations straightforward, as it enabled us to exploit the symmetry associated with the Killing vector $\xi$. It will certainly be important to examine how superradiance is affected by eccentricity and the emission of gravitational waves. That being said, \eqref{eq:A} is expected to be a good approximation for the amplification of wave packets that traverse the length of the (quasicircular) binary in a time much shorter than the orbital decay timescale ${t_\text{GW} \sim (\Omega \nu v^5)^{-1}}$ \cite{Blanchet:2013haa}.

Plugging in numbers showed that---at least within the EFT's regime of validity---the amplification factor is always small and therefore unobservable. However, the possibility remains that superradiance is more pronounced outside this regime. Systems in which ${ a/\lambda \sim \mathcal O(1) }$ are likely to be especially interesting, as resonant effects may occur. Recent numerical work has made progress in this direction \cite{VitorEtAl}, although it is worth exploring if the problem can also be studied analytically. Systems with ${ a \sim \lambda }$ can still have $v$ and ${ GM\w \ll 1 }$; thus, while zooming out on the binary as a whole is no longer a viable option, approximating the individual black holes by point particles remains a valid step.

Finally, we have focused solely on scattering in this work, but it is well known that superradiance can also manifest as an instability that triggers the exponential growth of bound states around Kerr black holes \cite{Detweiler:1980uk,Zouros:1979iw,Dolan:2007mj,Pani:2012bp,Witek:2012tr,Okawa:2014nda,Zilhao:2015tya,East:2017ovw,Brito:2013wya} and other astrophysical bodies \cite{Cardoso:2015zqa,Cardoso:2017kgn,Day:2019bbh}. A similar instability is likely to be present in black hole binaries and will be the subject of a forthcoming paper. In the future, it will also be worth extending all of these results to higher-spin fields.

\begin{acknowledgments}
It is a pleasure to thank Vitor Cardoso, Anne-Christine Davis, Josh Kirklin, Nakarin Lohitsiri, Harvey Reall, and Jorge Santos for helpful comments and discussions.
This work was partially supported by STFC Consolidated Grant No.~ST/P000681/1, and by a studentship from the Cambridge Commonwealth, European and International Trust.
\end{acknowledgments}

\appendix*
\section{}

This Appendix derives the result in \eqref{eq:A}. We begin by substituting \eqref{eq:q} into \eqref{eq:O} to obtain
\begin{equation}
O^L(t) = -\sum_\K \sum_{\ell'} \frac{A_\K}{\ell!\ell'!} \bmf z_\K^\avg{L}(t)
\frac{\dx}{\dx t} ( \bmf z_\K^\avg{L'}(t)\partial_{L'} \phi(t,\bmf 0) ).
\end{equation}
To proceed, we introduce the set of tensors $\Y^{\ell m}_L \equiv \Y^{\ell m}_{i_1 \dots i_\ell}$, which form basis sets for the vector spaces of STF tensors of rank $\ell$ \cite{Thorne:1980ru}. These objects generate the spherical harmonics:
\begin{equation}
\label{eq:Y_sph_harmonic}
Y_{\ell m}(\hat{\bmf x}) = \Y^{\ell m}_L \hat{\bmf x}^L;
\end{equation}
are orthogonal:
\begin{equation}
\label{eq:Y_orthogonal}
(\Y^{\ell m}_L)^* \Y^{\ell m'}_L = \frac{(2\ell+1)!!}{4\pi\ell!} \delta^{mm'};
\end{equation}
and satisfy the identity \cite{Endlich:2016jgc}
\eqskip
\begin{equation}
\label{eq:Y_rotation_property}
\Y^{\ell m}_{i_1 \dots i_\ell} R^{i_1}{}_{j_1}(t) \dots R^{i_\ell}{}_{j_\ell}(t) = \Y^{\ell m}_{j_1 \dots j_\ell} e^{im\Omega t}.
\end{equation}
These identities enable us to expand $\bmf z_\K^\avg{L}$ as
\begin{equation}
\bmf z_\K^\avg{L}(t) =  \frac{4\pi\ell!}{(2\ell+1)!!} r_\K^\ell
\sum_{m=-\ell}^\ell  Y^*_{\ell m}(\bmf d) \Y^{\ell m}_L e^{-im\Omega t},
\label{eq:id_1}
\end{equation}
such that we can write $O^L(t) = -4\pi i \sum_m \Y^{\ell m}_L O_{\ell m}(t)$, with
\begin{align}
O_{\ell m}(t)
&= \sum_{\ell',m'} \frac{4\pi Y^*_{\ell m}(\bmf d) Y_{\ell'm'}(\bmf d)}{(2\ell+1)!!(2\ell'+1)!!}
\sum_\K A_\K r_\K^{\ell+\ell'}
\nonumber\\
&\quad\times
e^{-i(m-m')\Omega t} (m'\Omega -i\partial_t) (\Y^{\ell'm'}_{L'})^*\partial_{L'}\phi(t,\bmf 0).
\label{eq:O_lm}
\end{align}

For the purposes of this paper, we want to evaluate the above using the zeroth-order solution
\begin{equation}
\phi^{(0)}(x) = \sum_{\ell,m}\int_\w e^{-i\w t} 2\mathcal I_{\w\ell m} j_\ell(kr) Y_{\ell m}(\hat{\bmf x}).
\end{equation}
Combining \eqref{eq:Y_sph_harmonic} with the fact that the spherical Bessel function ${ j_\ell(z) = [h_\ell^+(z) + h_\ell^-(z)]/2 }$ has the limiting form ${ j_\ell(z) \sim z^\ell/(2\ell+1)!! }$ as $z \to 0$, we find
\begin{equation}
\partial_L \phi^{(0)}(t,\bmf 0) =
\frac{ 2\ell! }{ (2\ell+1)!! }
\sum_m\int_\w\mathcal I_{\w\ell m}\Y^{\ell m}_L k^\ell e^{-i\w t}.
\label{eq:id_2}
\end{equation}
Now substitute this into \eqref{eq:O_lm} and use \eqref{eq:Y_orthogonal} to obtain
\begin{align}
O_{\ell m}(t)
&= \sum_{\ell',m'}\int_{\w'}
\frac{2 Y^*_{\ell m}(\bmf d) Y_{\ell'm'}(\bmf d) \mathcal I_{\w'\ell'm'}}{(2\ell+1)!!(2\ell'+1)!!}
\sum_\K A_\K r_\K^{\ell+\ell'}
\nonumber\\
&\quad\times
(k')^{\ell'}(m'\Omega -\w') e^{-i[\w' + (m-m')\Omega]t}.
\end{align}

As a final step, we have to show how \eqref{eq:phi_sol_1} reduces to \eqref{eq:phi_sol_1_A}. Performing the integral over $\bmf x'$ is a standard exercise  that yields~\cite{WDG}
\begin{equation}
\phi^{(1)}(x) = \sum_{\ell,m}\int_\w  (-1)^\ell\partial_L \int\dx t'
\Y^{\ell m}_L O_{\ell m}(t')\frac{e^{-i\w(t-t') + ikr}}{ir}.
\end{equation}
To simplify this further, a selection of identities \cite{Blanchet:1985sp,DLMF} can be used to show that
\begin{align}
\Y^{\ell m}_L \partial_L \frac{e^{ikr}}{ir}
&= \Y^{\ell m}_L \hat{\bmf x}^L r^\ell \left(\frac{1}{r}\frac{\dx}{\dx r}\right)^\ell k h_0^+(kr)
\nonumber\\
&= Y_{\ell m}(\hat{\bmf x}) k^{\ell+1} (-1)^\ell h_\ell^+(kr),
\end{align}
and therefore
\begin{equation}
\mathcal A_{\w\ell m} =
k^{\ell+1} \int\dx t' O_{\ell m}(t') e^{i\w t'}.
\end{equation}
Completing the final integral over $t'$, which trivially yields a delta function, gives us the desired result in \eqref{eq:A}.

\bibliography{sup}
\end{document}